# Vortex channeling and the *V-I* characteristic of YBa$_2$Cu$_3$O$_7$ low angle grain boundaries

M. J. Hogg, F. Kahlmann, E. J. Tarte, Z. H. Barber, and J. E. Evetts

Department of Materials Science and IRC in Superconductivity, University of

Cambridge, Pembroke Street, Cambridge CB2 3QZ, United Kingdom

We have performed *V-I* measurements on a thin film YBa$_2$Cu$_3$O$_7$ 4° [001] tilt low angle grain boundary over an extensive range of temperature and field, verifying the presence of a linear characteristic. We report on the occurrence of the linear characteristic in its basic form and on the observation of *V-I* kinking into several, and in some cases numerous, linear segments. We interpret these findings in terms of a variation in the dissipative width at the grain boundary. Kinking from one linear *V-I* section to another of different gradient is described in terms of a change in the number of vortex rows being viscously channeled along the boundary.

In biaxially textured $YBa_2Cu_3O_7$ (YBCO) "coated conductor" tapes the critical current density ($J_c$) is now limited by networks of low angle grain boundaries[1] (LAGBs). As a consequence, LAGBs have become a natural focus for investigation, particularly in terms of the elucidation of their electromagnetic behavior. In general, YBCO [001] tilt LAGBs consist of relatively undisturbed lattice[2,3] interspersed by regularly spaced edge dislocations around which strain fields[4] and band structure effects[5] locally suppress superconductivity. The $J_c$ of a current track containing a LAGB is determined by the balance of two opposing effects, (1) the localized reduction in current carrying cross section[2] and (2) the enhanced flux pinning due to accommodation of vortices to the dislocation array[6]. There is a lack of consensus on the shape of the LAGB *V-I* characteristic above $J_c$ and the dissipation mechanism that determines it. Smaller angle LAGBs are reported to have a curved characteristic[7], referred to as 'flux flow', and larger angles[8] (10°) have been shown to exhibit RSJ-like negative curvature transitions. However, measurements made by Díaz *et al.*[9] on an isolated 4° boundary found the characteristic to be linear for certain fields and temperatures, a result recently confirmed by Verebelyi *et al.*[7] for a 4.5° misorientation LAGB. We believe the linear LAGB characteristic is indicative of dissipation dominated by *viscous flux flow* (VFF). VFF is expected to dominate over flux creep effects since localized *E* fields are very large (on the order $10^2$V/m)[9].

Another possible source of LAGB *V-I* linearity is long Josephson junction behavior, a characteristic of high angle grain boundaries. Although this cannot be ruled out as a contributory factor, the observation[8] of a strong-coupled to Josephson-coupled crossover in a 10° junction at temperatures above 75K tends to support the VFF model for lower misorientations at comparable or lower temperatures.

We have performed *V-I* measurements on a thin film YBCO bicrystal LAGB over a range of temperature (10K to 85K) and field (0T to 8T). In addition to the occurrence of linear transitions we report the observation of kinking, or faceting, of the curve into several, and in some cases numerous linear segments (see Fig. 1). We interpret these findings in terms of the variation of the dissipative width ($d_{eff}$) at the interface.

An YBCO film was grown on a nominally 4° [001] tilt SrTiO$_3$ bicrystal by off-axis ablation at a temperature of 790°C and annealed for 20 minutes at 500°C in 1 atm O$_2$. The film had a $T_c$ of 89.5K. X-ray measurements resulted in a rocking curve FWHM of 0.28° and were used to determine the precise grain boundary (GB) misorientation to be 3.8° [001] tilt (±0.5°) and less than 0.5° [100] tilt and [100] twist. Tracks were patterned both across the LAGB and within the adjacent epitaxial grain using standard photolithographic techniques to allow four point electrical measurements (using a Keithley 182 voltmeter and an HP 3245A current source). AFM measurements were used to confirm that the film was 150nm thick and to reveal that the YBCO GB followed the underlying interface with a meander of less than 100nm. Significantly larger deviations and holes were, however, observed with an average frequency of once per micron and almost certainly correspond to large hole defects (~100nm diameter, 10-20nm deep) imaged in the underlying substrate interface. Such defects are commonly found in SrTiO$_3$ bicrystal substrates. They have been found to deleteriously influence the transport properties high angle boundaries[10] and may modulate $d_{eff}$ along the YBCO interface in LAGBs. TEM studies have also shown the YBCO boundary to be faceted in high angle thin film[11] and low angle flux grown (6°)[12] samples, leading to strain fields extending up to 30nm into the grains.

Linearity is observed in various forms in most of our *V-I* transitions, ranging from a classic single linear trace found at higher temperatures to piecewise linear traces more frequently encountered at lower temperatures. A particularly clear example of kinking is shown in the inset Fig. 1 for a LAGB track of width 9.1μm. Since *E-J* requires a precise knowledge of $d_{eff}$, we plot *V-J* where *J=I/A* (*A* being track cross section). The transition above $J_c$ can be seen to comprise two linear sections of differing gradient, d*V*/d*J* becoming discontinuous at the kink between segments.

Examples of *V-J* temperature dependence are shown in Fig. 1. At 75K, *V-J*s show clear linearity up to a dissipation level of 10µV. As *T* is reduced, both an increase in $J_c$ and a decrease in d*V*/d*J* (due to a corresponding reduction in $\rho_f$ with *T*) are observed. Further reduction in *T* results in the appearance of multiply kinked *V-J* transitions, such as the example shown at 35K. These findings are in marked contrast to the smoothly curved intragranular (IG) transitions measured in the grains over the same wide range of *B* and *T*, also plotted in Fig 1 for comparison. *Smoothly* curved LAGB *V-J*s are only observed at large applied fields when the granular irreversibility field is exceeded.

In order to explain the occurrence of kinking we use a modification of a simple flux-channeling model recently proposed by Evetts *et al.*[13]. They modeled the *V-I* characteristic of a coated conductor assuming a 2-D network of LAGBs and the formation of percolative vortex channels along GB segments. By assuming a constant $d_{\text{eff}}$ (one vortex row), the constraint that ∆*V* across each vortex path be constant along the entire path length necessitates a constant LAGB *E* field and vortex flow velocity, $v_f$, through the relation $\mathbf{E} = -\mathbf{v}_f \times \mathbf{B}$. Whereas for a granular coated conductor a kink in the sample *V-I* is predicted to occur due to the formation of an additional vortex percolation channel, in the case of a single LAGB we can attribute kinking to a change in the number of vortex rows flowing at the LAGB. We propose that this variation in dissipative width reflects the presence of an effective $J_c$ profile across the boundary (Fig. 2). Moving away from the LAGB, $J_c$ gradually increases from a minimum value ($J_c = J_c^{GB}$) within the interface to a maximum given by the IG critical current density ($J_c = J_c^{IG}$). The effects of such a profile will depend upon its relative width with respect to the local vortex spacing (function of *B*) and size (function of *T*).

If we assume that each row flows with a linear *V-I* given by a constant flux flow resistivity, $\rho_f$, then for the *k*th row $E_k = \rho_f [J(k) - J_c(k)]$. If the width of one row is $\sqrt{3}a_o/2$, then the voltage drop across the track will be a summation of $\Delta V_k = (\sqrt{3}a_o/2)E_k$ over $m$ flowing vortex rows

$$V = \sum_{k=1}^{m} \Delta V_k = \sum_{k=1}^{m} \frac{\sqrt{3}}{2} a_o E_k = \sum_{k=1}^{m} \frac{\sqrt{3}}{2} a_o \rho_f [J(k) - J_c(k)] \quad (1)$$

whence the slope of the $V$-$J$ characteristic will become

$$\frac{dV}{dJ} = \gamma_f m(J) \qquad \text{where} \qquad \gamma_f = \frac{\sqrt{3}}{2} a_o \rho_f \quad (2)$$

This simple result predicts that the number of active vortex rows, $m = m(J)$ a function of the overall current density, will determine the gradient of the $V$-$J$ transition as an integer multiple of $\gamma_f$ (assumed constant for a given $B$ and $T$). Thus, as illustrated in Fig. 3, a change in the number of rows flowing will result in a step increment in the track $dV/dJ$.

Although the characteristics we observe do point to a VFF mechanism it is clear that the full description is somewhat more complex; we find that the increment in gradient is not always in integer, or sequentially increasing, steps of $\gamma_f$. The position of kinks and sections of linearity are frequently irreproducible from scan to scan and are often absent in some $V$-$J$s at lower temperatures, which exhibit instead a noisy erratic curvature. Furthermore, the $V$-$J$ curve itself (not just $dV/dJ$) is commonly observed to become discontinuous at kink positions and is accompanied by an extremely pronounced $B$ and $T$ dependent noise level. Examples of such transitions, containing segments of linearity, but also characterized by noise on various $J$ scales, are shown in Fig. 4(a). IG transitions are plotted in Fig. 4(b) for comparison. Note that noise only appears above $J_c$ and does not appear in the IG transitions, suggesting that it is intrinsic to the LAGB and not indicative of instrumental effects.

These additional features can be explained by relaxing the simplifying assumptions made in our treatment. In practice the LAGB is inhomogeneous on several length scales and there will be deviations in size and shape of the $J_c$ profile along the length of the interface. Such deviations may give rise to the channeling of partial rows of vortices, and indeed, a simple modification to the above treatment shows that non-integer steps of $\gamma_f$ would be expected for such behavior[14]. Variation in vortex viscosity (due to variation in vortex core size at the LAGB) could also lead to such an effect. Multiple non-integer kinking of the *V-J* transition could, as a result, produce a characteristic with a more curved nature. *V-J* discontinuities would be expected if the additional rows of vortices starting to flow affected the properties, in particular $J_c$ and $\gamma_f$, of rows already flowing. Interaction between flowing vortex rows and the static IG vortex configuration may lead to a modified flow pattern so as to minimize entropy production. It is expected that multiple kinking in d*V*/d*J* combined with *V-J* discontinuities will result transition noise.

We have presented a flux flow based model that explains key elements of *V-J* transitions measured for a 4° LAGB. Compared to granular *V-J*s the linear, kinked and featured LAGB characteristics support to various extents our interpretation of the dissipative mechanism in terms of VFF. It is accepted, however, that long junction effects may play an increasingly significant role as the angle of misorientation of the LAGB is increased. In order to clarify the range of applicability a more extensive analysis is needed involving scatter plots of extracted $\rho_f$ values as a function of *B* and *T*.

Support is acknowledged from the EPSRC and the EC under TMR Network No. CT98-0189 SUPERCURRENT.

# Figure Captions

**FIG. 1.** *V-J* measurements taken at 1T for a 5.5µm wide, 4.6° LAGB track at 35K and 75K. An IG transition, also taken at 35K 1T, is plotted on the upper scale for comparison. The inset shows a particularly clear example of kinking between two linear segments taken at 1.9T, 60K for a 9.1µm wide LAGB track. Dashed lines are a guide to the eye.

**FIG. 2.** Proposed effective $J_c$ profile (of width *x*) across an LAGB, leading to variation in dissipative width as the number of vortex rows channeled along the boundary changes with transported current density, where $d_{\text{eff}} = m(J)\sqrt{3}a_o/2$. Vortices will be initially depinned at the LAGB, surrounding rows progressively depinning as *J* is increased.

**FIG. 3.** Kinked LAGB *V-J* expected for an assumed *m*(*J*). In this case *m* increases by one in regular intervals of *J*. The $k^{\text{th}}$ vortex row starts to flow when $J = J_c(k)$, where $J_c(k) < J_c(k+1)$ and $J_c^{\text{GB}} = J_c(1)$. The track *V-J* (thick solid line) is simply a summation of individual vortex row *V-J*s (thin solid lines), a step increment in d*V*/d*J* (marked by arrows) being associated with a change in the number of rows flowing.

**FIG. 4.** *V-J* characteristics measured for (a) LAGB and (b) IG tracks at 50K in fields of 0.4T, 0.6T, 0.8T, 1.5T and 5T. The LAGB transitions show segments of **linearity and kinking, but are also characterized to various extents by degrees of noise. In marked contrast, the IG measurements display smoothly curved transitions at the same dissipation (voltage) levels.**

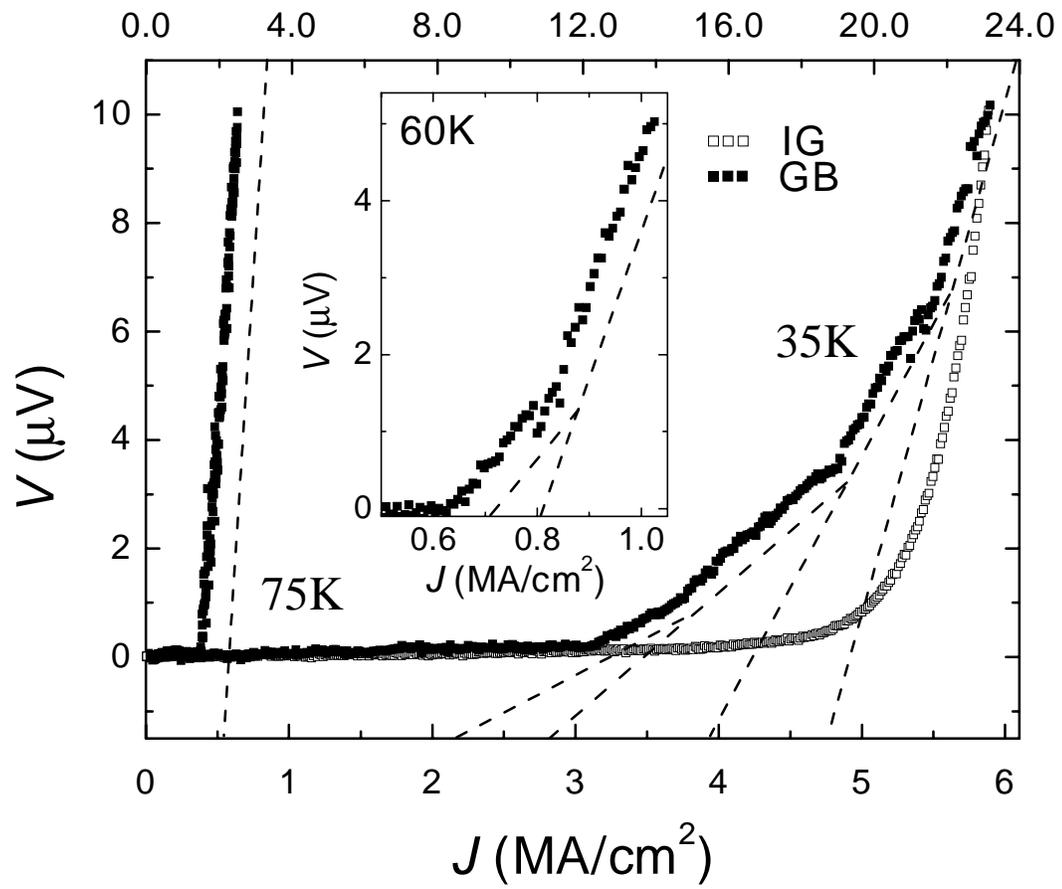

Fig. 1.
(Hogg *et al.*)

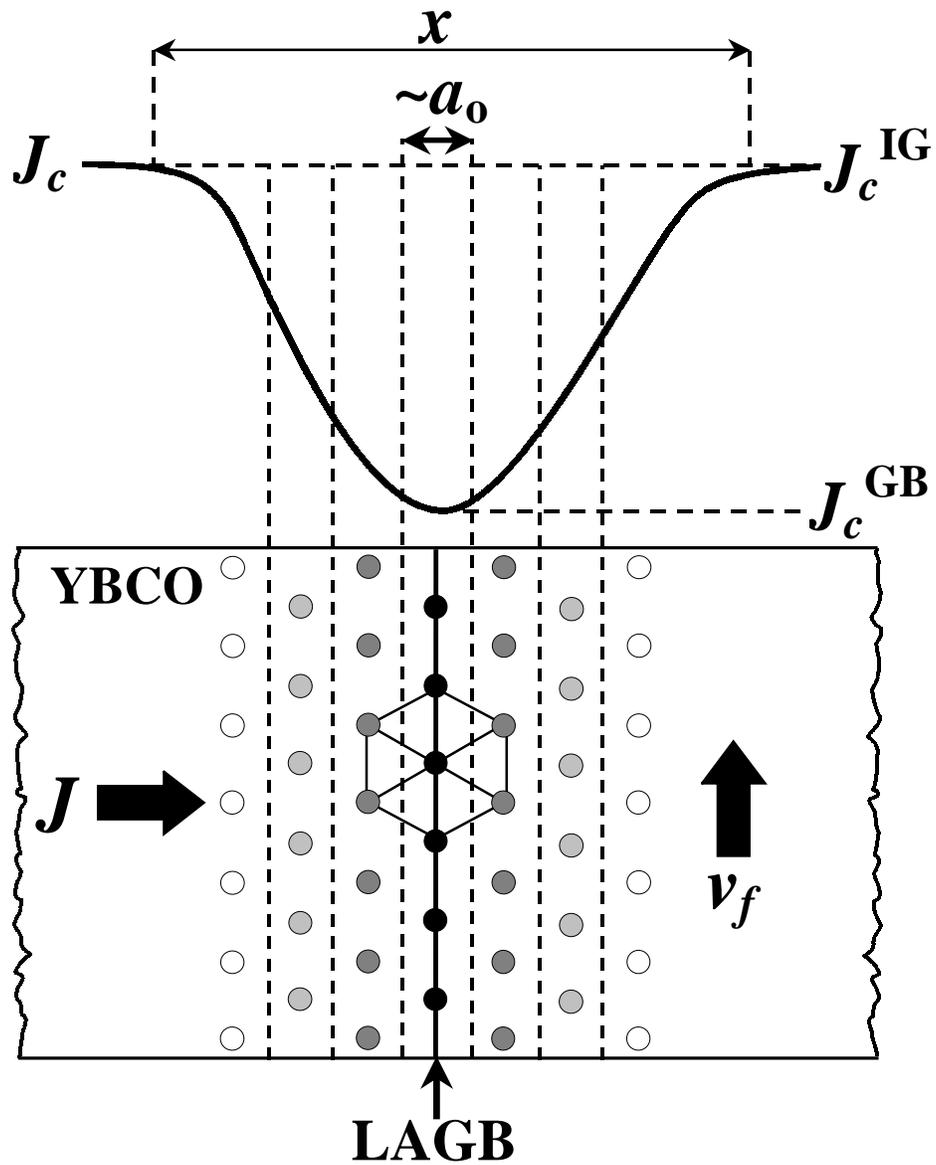

Fig. 2.
(Hogg *et al.*)

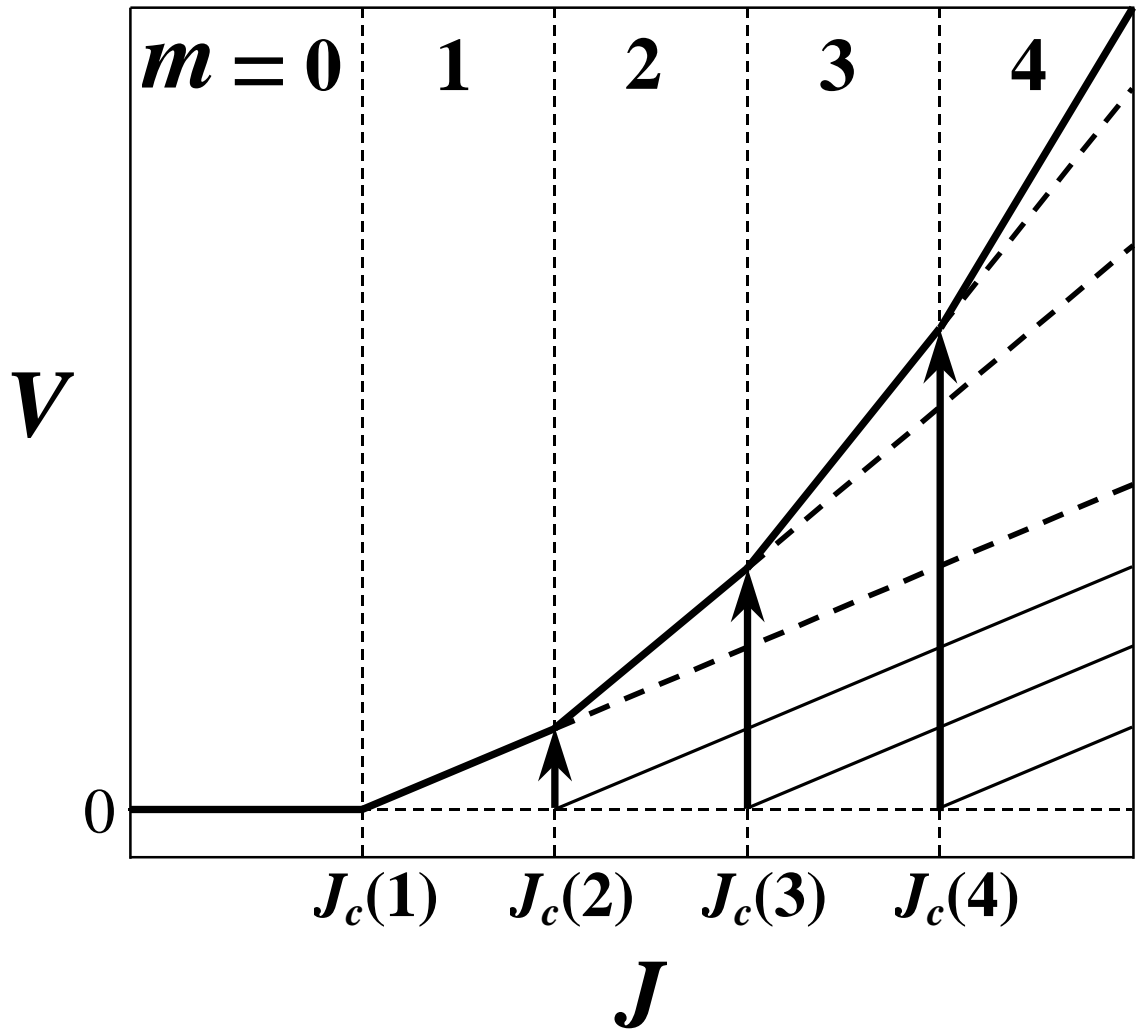

Fig. 3
(Hogg *et al.*)

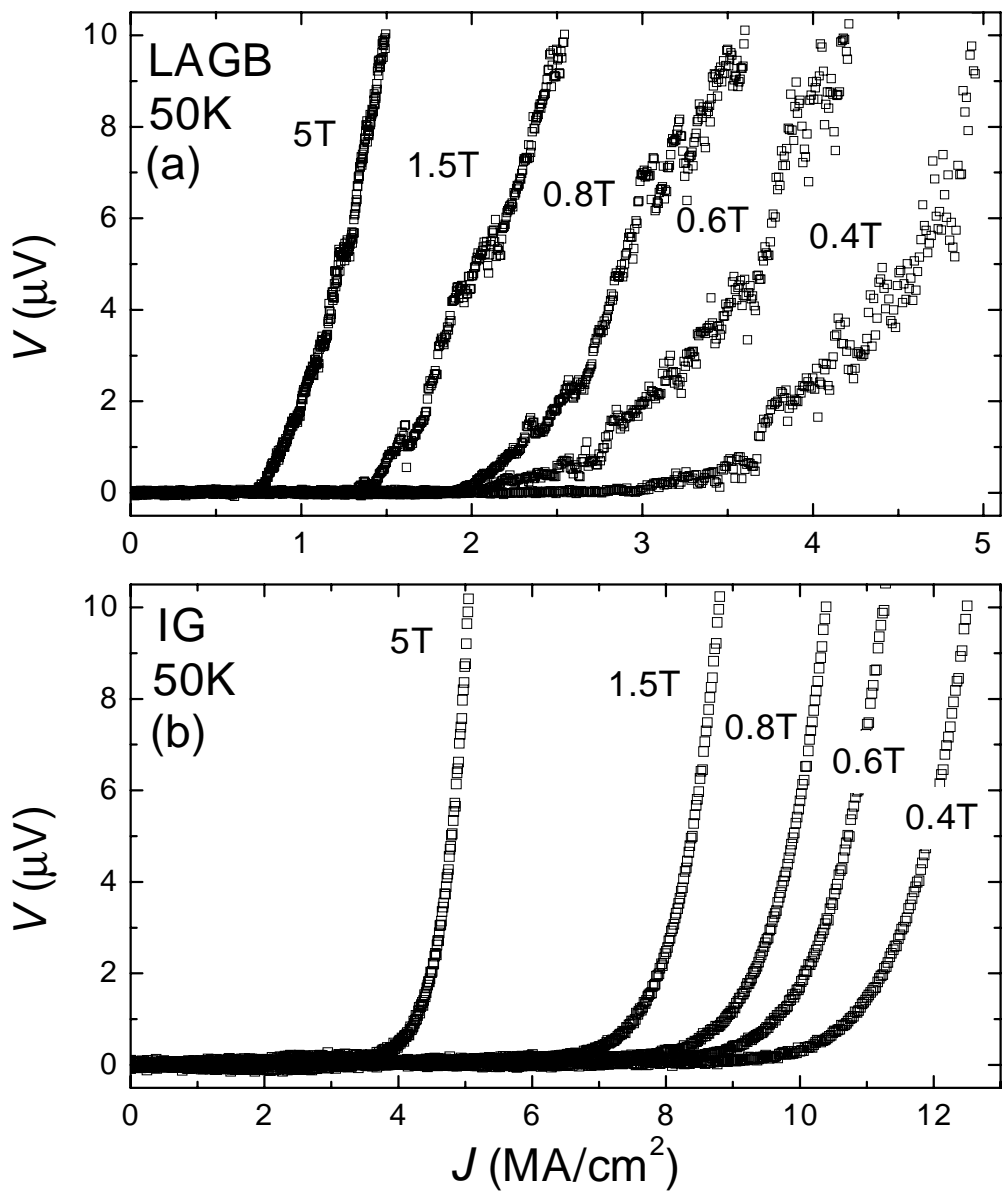

Fig. 4.
(Hogg *et al.*)